\documentclass[11pt]{article}
\usepackage{epsfig,cite}
\usepackage{amsmath}
\usepackage{amssymb}
\usepackage{amsfonts}
\usepackage{latexsym}
\usepackage{wasysym}
\usepackage{url}
\usepackage{hyperref}
\usepackage{bm}
\usepackage{graphics}
\usepackage{rotating}
\usepackage{booktabs}

\def\bea{\begin{eqnarray}}
\def\eea{\end{eqnarray}}
\def\beq{\begin{equation}}
\def\eeq{\end{equation}}
\catcode`@=11 
\def\slash#1{\mathord{\mathpalette\c@ncel#1}}
 \def\c@ncel#1#2{\ooalign{$\hfil#1\mkern1mu/\hfil$\crcr$#1#2$}}
\def\lsim{\mathrel{\mathpalette\@versim<}}
\def\gsim{\mathrel{\mathpalette\@versim>}}
 \def\@versim#1#2{\lower0.2ex\vbox{\baselineskip\z@skip\lineskip\z@skip
       \lineskiplimit\z@\ialign{$\m@th#1\hfil##$\crcr#2\crcr\sim\crcr}}}
\catcode`@=12 

\def\({\left(}
\def\){\right)}
\def\[{\left[}
\def\]{\right]}

\relax

\def    \hepph  #1 {{\tt hep-ph/#1}}
\def    \hepex  #1 {{\tt hep-ex/#1}}

\newcommand{\be}{\begin{equation}}
\newcommand{\ee}{\end{equation}}

\begin{document}
\begin{figure}[h]
\epsfig{width=0.32\textwidth,figure=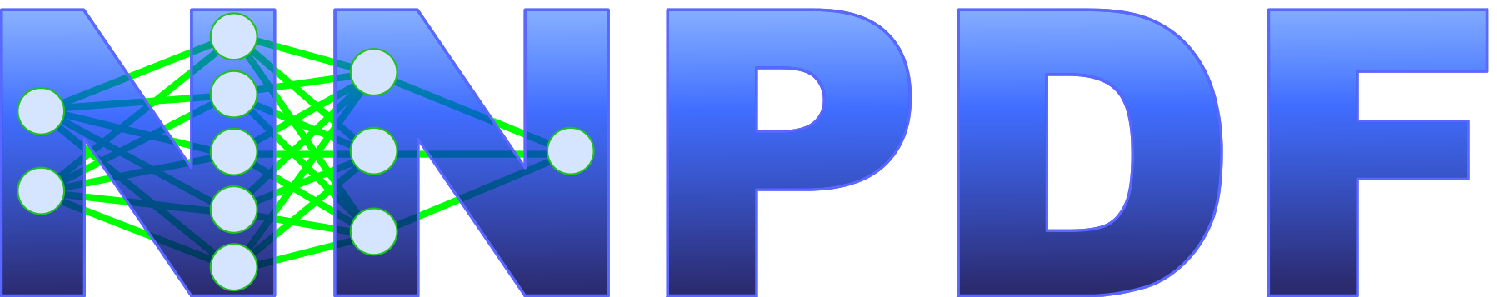}
\end{figure}
\vspace{-2.0cm}

\begin{flushright}
\ \\ \ \\ \ \\ 
CERN-PH-TH/2013-210 \\
Edinburgh 2013/26 \\
IFUM-1016-FT\\
\end{flushright}
\begin{center}
{\Large \bf Polarized Parton Distributions at an 
Electron-Ion Collider}
\vspace{0.8cm}

{\bf The NNPDF Collaboration}: Richard~D.~Ball,$^{1}$ Stefano~Forte,$^2$
Alberto~Guffanti,$^{3}$\\  
Emanuele~R.~Nocera,$^2$  Giovanni~Ridolfi\,$^4$ and
Juan~Rojo.$^5$

\vspace{1.cm}

{\it ~$^1$ Tait Institute, University of Edinburgh,\\
JCMB, KB, Mayfield Rd, Edinburgh EH9 3JZ, Scotland\\
~$^2$ Dipartimento di Fisica, Universit\`a di Milano and
INFN, Sezione di Milano,\\ Via Celoria 16, I-20133 Milano, Italy\\
~$^3$ The Niels Bohr International Academy and Discovery Center, \\
The Niels Bohr Institute, Blegdamsvej 17, DK-2100 Copenhagen, Denmark\\
~$^4$ Dipartimento di Fisica, Universit\`a di Genova and
INFN, Sezione di Genova,\\ Genova, Italy\\
~$^5$ PH Department, TH Unit, CERN, CH-1211 Geneva 23, Switzerland \\
}

\end{center}

\vspace{0.8cm}

\begin{center}
{\bf \large Abstract:}
\end{center}
We study the potential impact of inclusive deep-inelastic scattering
data
from a future 
electron-ion collider (EIC) on longitudinally polarized  parton
distributions (PDFs).
We perform a PDF 
determination using the NNPDF methodology, 
based on sets of deep-inelastic EIC pseudodata, for
different realistic choices of
the electron and proton beam energies.
We compare the results to our current polarized PDF set,
\texttt{NNPDFpol1.0}, based on a fit to fixed-target
inclusive DIS data. 
We show that the uncertainties on the first moments of the polarized quark
singlet and gluon 
distributions are substantially reduced in comparison to
\texttt{NNPDFpol1.0}, but also that more measurements may be needed 
to ultimately pin down
the size of the gluon contribution to the nucleon spin.

\clearpage

The accurate determination of polarized parton distribution functions (PDFs),
along with a reliable estimate of their uncertainties, has been the goal of several recent 
studies~\cite{deFlorian:2009vb,Leader:2010rb,Hirai:2008aj,Blumlein:2010rn,Ball:2013lla}.
In Ref.~\cite{Ball:2013lla} we presented a first determination of polarized PDFs based on
the NNPDF methodology~\cite{Ball:2013lla}, {\tt NNPDFpol1.0}, which uses Monte Carlo 
sampling for error propagation and unbiased PDF parametrization in terms of neural networks.
The bulk of experimental information on longitudinally polarized proton 
structure comes from inclusive neutral-current deep-inelastic scattering (DIS),
which allows one to obtain information on the light quark-antiquark combinations
$\Delta u +\Delta\bar{u}$, $\Delta d +\Delta\bar{d}$,  
$\Delta s +\Delta\bar{s}$ and on the gluon distribution $\Delta g$.
However, DIS data covers only a small kinematic region of momentum fractions and 
energies $(x,Q^2)$.
On the one hand, the lack of experimental information for $x\lesssim10^{-3}$ prevents a 
reliable determination of polarized PDFs at small-$x$. 
Hence, their first moments
will strongly depend on the functional form one assumes
for PDF extrapolation to the unmeasured $x$ region~\cite{deFlorian:2009vb}.  
On the other hand, the gluon PDF, which is determined 
by scaling violations, is only 
weakly constrained, due to the small lever-arm in $Q^2$
of the experimental data. 

For these reasons, despite many efforts, both experimental and theoretical, 
the size of the polarized  gluon contribution to the nucleon spin 
is still largely uncertain~\cite{deFlorian:2011ia,Ball:2013lla}.
Open charm photoproduction data from COMPASS~\cite{Adolph:2012ca} do not
change this state  
of affairs: they were  shown in Ref.~\cite{Nocera:2013yia} to have  
almost no impact on $\Delta g$.
Present and future polarized hadron collider measurements 
from RHIC~\cite{Adare:2008qb,Adamczyk:2012qj,Adamczyk:2013yvv,Adare:2008aa,Adare:2010cc}, 
specifically semi-inclusive particle 
production and jet data, should provide further constraints
on $\Delta g$, but restricted to the medium- and large-$x$ region.

An Electron-Ion Collider (EIC)~\cite{Deshpande:2005wd,Boer:2011fh,Accardi:2012hwp}, with
polarized lepton and hadron beams, would allow for a widening of the
kinematic region comparable to the one achieved in the unpolarized case with the
DESY-HERA experiments H1 and ZEUS~\cite{Aaron:2009aa}
(note that a Large Hadron-electron Collider 
(LHeC)~\cite{AbelleiraFernandez:2012cc} would not have the option of
polarizing the hadron beam).
 The potential impact of the EIC on the knowledge of the nucleon spin 
has been quantitatively assessed
in a recent study~\cite{Aschenauer:2012ve}, in which 
projected neutral-current inclusive DIS and semi-inclusive DIS (SIDIS) 
artificial data were added to the DSSV polarized PDF
determination~\cite{deFlorian:2011ia}; this study was then  extended 
by also providing an estimate of the impact of charged-current inclusive DIS pseudo-data on the 
polarized quark flavor separation in Ref.~\cite{Aschenauer:2013iia}.
In view of the fact that a substantially larger gluon uncertainty was
found in Ref.~\cite{Ball:2013lla} in comparison to previous PDF
determinations~\cite{deFlorian:2009vb,Leader:2010rb,Hirai:2008aj,Blumlein:2010rn}, 
it is worth repeating the study of the impact of EIC
data, but now using NNPDF methodology. 
This is the goal of the present paper.

Two alternative designs have been proposed for the EIC so far: 
the electron Relativistic Heavy Ion Collider (eRHIC) at  
Brookhaven National Laboratory (BNL)~\cite{BNL:eRHIC} 
and the Electron Light Ion Collider (ELIC)
at Jefferson Laboratory (JLab)~\cite{JLAB:ELIC}.
In both cases, a staged upgrade of the existing facilities
has been planned~\cite{Deshpande:2005wd,Boer:2011fh,Accardi:2012hwp}, so that
an increased center-of-mass energy would be available at each stage.
Concerning the eRHIC option of an EIC~\cite{BNL:eRHIC},
first measurements would be taken by colliding the 
present RHIC proton beam of energy $E_p=100-250$~GeV
with an electron beam of energy $E_e=5$ GeV,
while a later stage envisages electron beams with energy up to $E_e=20$ GeV.

In order to quantitatively assess the impact of the EIC data, 
we have supplemented our previous QCD analysis~\cite{Ball:2013lla}
with  DIS pseudodata from Ref.~\cite{Aschenauer:2012ve},
which consist of three sets of data points at different possible
eRHIC electron and proton beam energies, as discussed above. 
These pseudodata were produced by running
the {\tt PEPSI} Monte Carlo (MC) generator~\cite{Mankiewicz:1991dp}, assuming
momentum transfer $Q^2>1$ GeV$^2$, squared invariant mass 
of the virtual photon-proton system $W^2>10$ GeV$^2$ and fractional energy
of the virtual photon $0.01\leq y \leq 0.95$; they are provided
in five (four)  bins per logarithmic decade in $x$ ($Q^2$). For each dataset,
the $Q^2$ range spans the values from $Q^2_{min}=1.39$ GeV$^2$ to 
$Q^2_{\rm max}=781.2$ GeV$^2$, while the accessible values of momentum
fraction $x=Q^2/(sy)$ depend on the available center-of-mass energy, 
$\sqrt{s}$.
In Tab.~\ref{tab:kintab}, we summarize, for each data set, the number of 
pseudodata $N_{\mathrm{dat}}$; the electron and proton beam 
energies $E_e$, $E_p$; the corresponding center-of-mass energies $\sqrt{s}$;
and the smallest and largest accessible value in the momentum fraction range,
$x_{\rm min}$  and $x_{\rm max}$ respectively.

\begin{table}[t]
\small
\centering
\begin{tabular}{llcccccc}
\toprule
Experiment& Set& $N_{\mathrm{dat}}$& 
$E_e\times E_p$ [GeV] & $\sqrt{s}$ [GeV]&
$x_{\rm min}$ & $x_{\rm max}$ & $\langle \delta g_1 \rangle$\\ 
\midrule
EIC 
& EIC-G1P-1 & 56 & $5\times 100$ & $44.7$ & 
$8.2\times 10^{-4}$ & $0.51$& $0.010$\\
& EIC-G1P-2 & 63 & $5\times 250$ & $70.7$ & 
$3.2\times 10^{-4}$ & $0.51$ & $0.032$ \\
& EIC-G1P-3 & 61 & $20\times 250$ & $141$ & 
$8.2\times 10^{-5}$ & $0.32$ & $0.042$\\
\bottomrule
\end{tabular}
\caption{\small The three  EIC pseudodata sets~\cite{Aschenauer:2012ve}. 
For each set we show the number of points
$N_{\mathrm{dat}}$, the electron and proton beam energies $E_e$  and $E_p$,
the center-of-mass energy $\sqrt{s}$, the kinematic coverage
in the momentum fraction $x$, and  the average absolute statistical
uncertainty $\langle \delta g_1 \rangle$.}
\label{tab:kintab}
\end{table} 

\begin{figure}[t]
\begin{center}
\epsfig{width=0.40\textwidth,figure=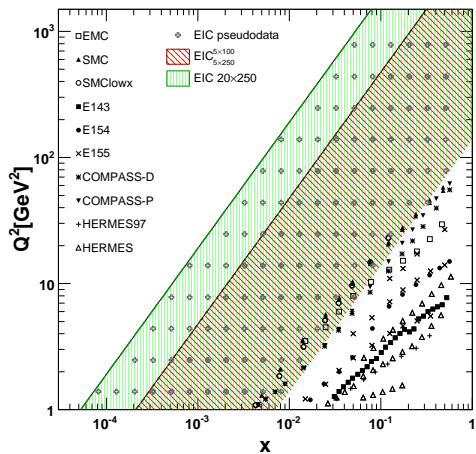}
\caption{\small Kinematic coverage in the $(x,Q^2)$ plane
for the fixed-target experimental data included in the {\tt NNPDFpol1.0} 
polarized parton fit~\cite{Ball:2013lla} 
and the  EIC pseudodata from~\cite{Aschenauer:2012ve}.
The shaded bands show the expected kinematic reach of
each of the two EIC scenarios discussed in the text.}
\label{fig:kinplot}
\end{center}
\end{figure}

The kinematic coverage of the EIC pseudodata
 is displayed in Fig.~\ref{fig:kinplot} together with
the fixed-target DIS data points included in our previous
analysis~\cite{Ball:2013lla}. 
The dashed regions show the overall kinematic reach of the EIC data
with the two electron beam energies $E_e=5$ GeV or
$E_e=20$ GeV, corresponding 
to  each of the two stages at eRHIC.
It is apparent from Fig.~\ref{fig:kinplot} that EIC data
will extend the kinematic coverage significantly, even for the lowest
center-of-mass energy. 
In particular, hitherto unreachable  small $x$ values, down to $10^{-4}$,
will be attained, thereby leading to a significant reduction of the 
uncertainty in the low-$x$ extrapolation region.
Furthermore,  the increased lever-arm
in $Q^2$, for almost all values of $x$, should allow for much more stringent
constraints on  $\Delta g(x,Q^2)$ from scaling 
violations.      

The observable provided in Ref.~\cite{Aschenauer:2012ve} for
inclusive DIS pseudodata is the ratio $g_1(x,Q^2)/F_1(x,Q^2)$; we refer
the reader to Ref.~\cite{Ball:2013lla} for a discussion of its relation to
experimentally measured asymmetries. The generation of pseudodata
assumes a ``true'' underlying set of parton distributions. In
Ref.~\cite{Aschenauer:2012ve} these are taken to be
{\tt DSSV+}~\cite{deFlorian:2011ia} and {\tt MRST}~\cite{Martin:2002aw} 
polarized and unpolarized PDFs 
respectively.
 Uncertainties are then determined assuming an integrated
luminosity of $10$ fb$^{-1}$, which 
corresponds to a few months operations for the anticipated
luminosities for eRHIC~\cite{BNL:eRHIC}, and a $70\%$ beam
polarization.
Because the {\tt DSSV+} polarized gluon has rather more structure than that
of {\tt NNPDFpol1.0}, which is largely compatible with zero,
assuming this input shape will allow us to test whether the EIC data
are sufficiently accurate to determine the shape of the gluon
distribution.

We reconstruct the $g_1$ polarized structure function from the pseudodata
following the same procedure used in Ref.~\cite{Ball:2013lla} for the
E155 experiment. We provide its average statistical uncertainty in
the last column of Tab.~\ref{tab:kintab}. A comparison 
of these values with the analogous quantities for 
fixed-target experiments (see Tab.~2 in Ref.~\cite{Ball:2013lla})
clearly shows that EIC data are expected to be far more precise, 
with uncertainties reduced up to one order of magnitude.
No information on the expected systematic uncertainties is available.
We will perform two different fits, corresponding to the two stages envisaged
for the eRHIC option of an EIC~\cite{BNL:eRHIC} discussed above, 
which will be referred to as \texttt{NNPDFpolEIC-A}
and \texttt{NNPDFpolEIC-B}. 
The former includes the first two sets of pseudodata listed in 
Tab.~\ref{tab:kintab}, while the latter also includes the third set.

\begin{figure}[h]
\begin{center}
\large{Distribution of $\chi^{2(k)}_{\mathrm{tot}}$ for individual sets}\\
\epsfig{width=0.40\textwidth, figure=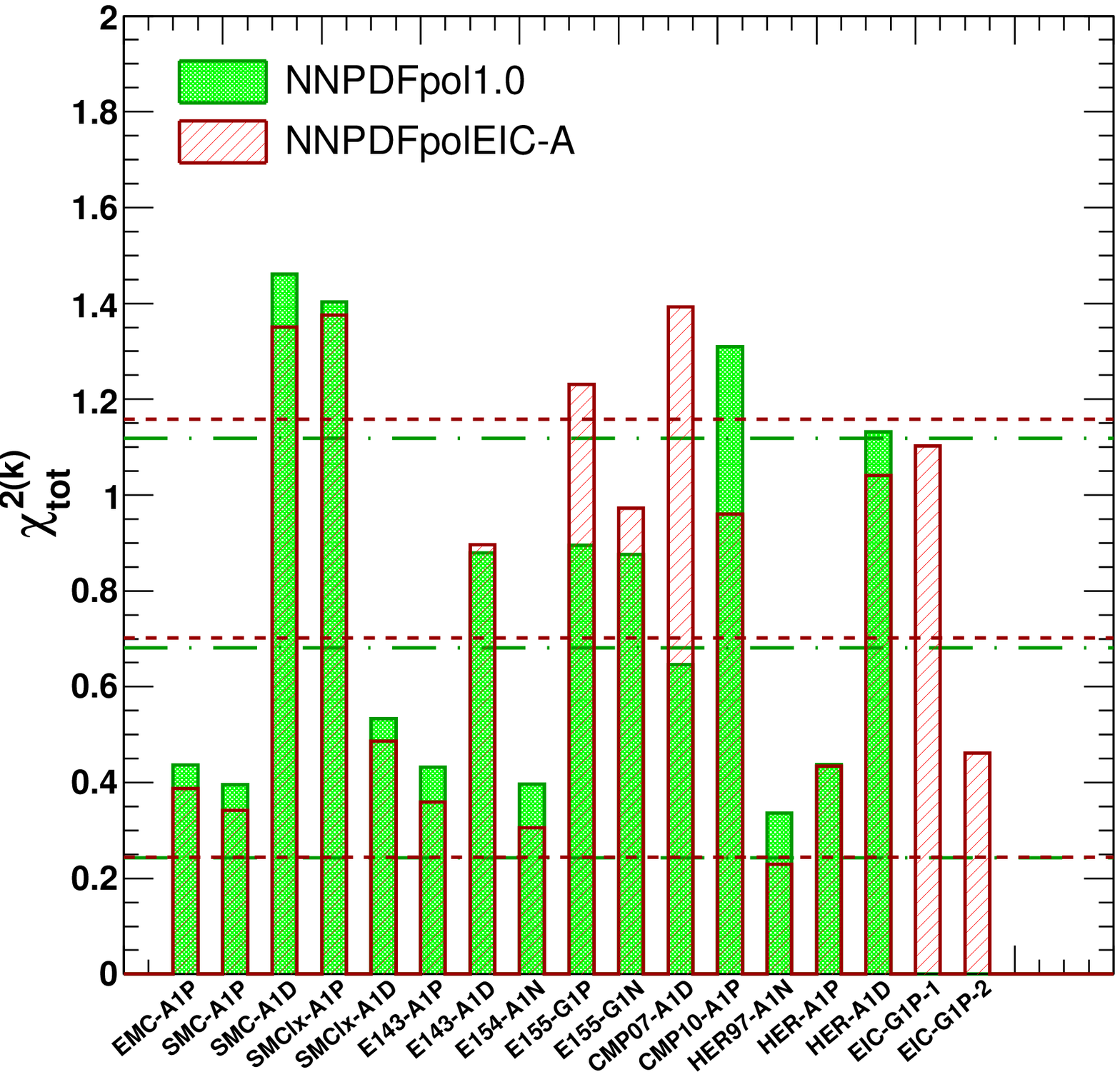}
\epsfig{width=0.40\textwidth, figure=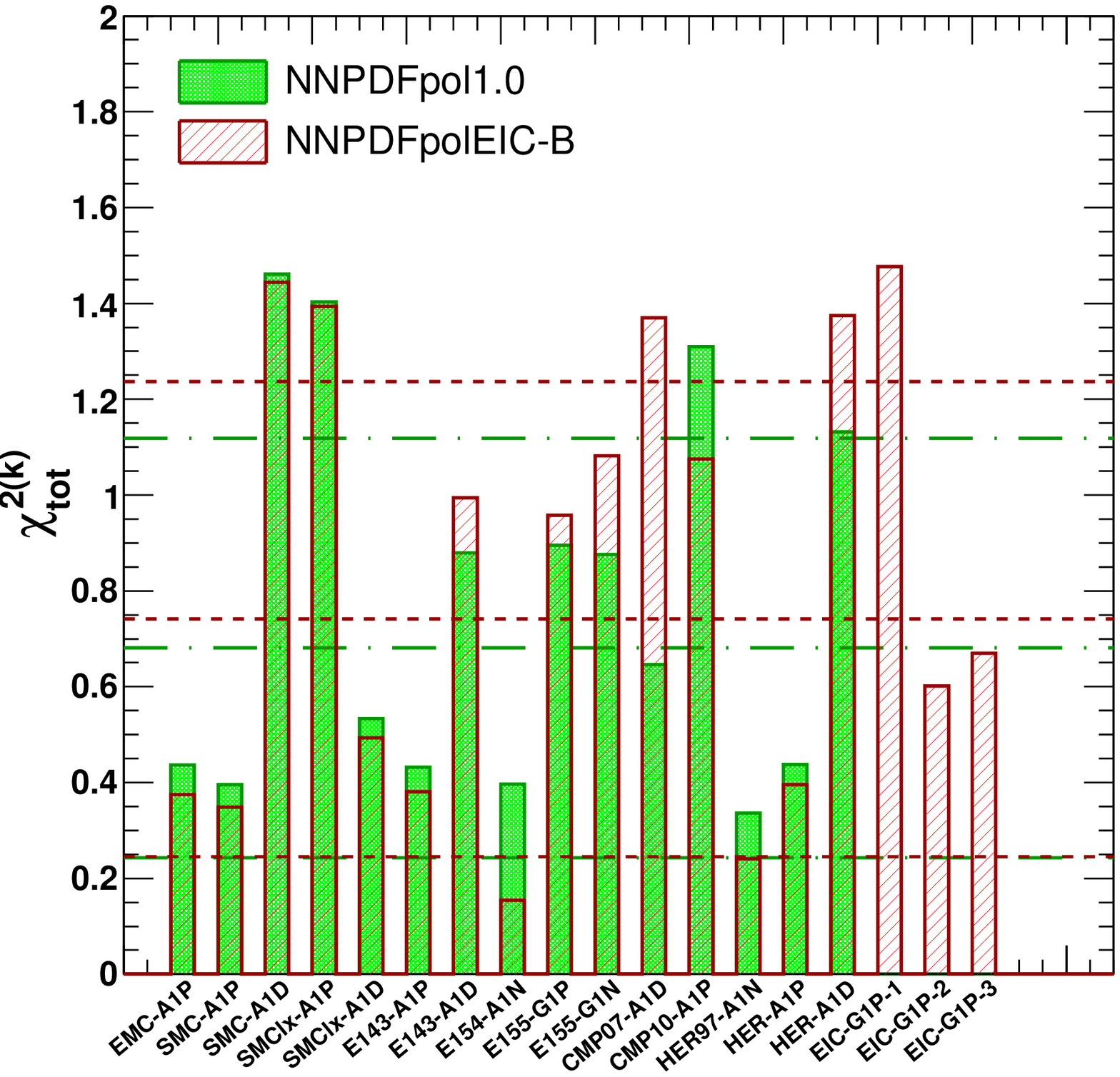}\\
\end{center}
\caption{Value of the $\chi^2$ per data point for the datasets 
included in the \texttt{NNPDFpolEIC-A} (left)
and in the \texttt{NNPDFpolEIC-B} (right) fits, 
compared to \texttt{NNPDFpol1.0}~\cite{Ball:2013lla}.
The horizontal lines correspond to the unweighted average of the 
$\chi^2$ values shown, and the one-sigma interval about it.
The dashed lines refer to \texttt{NNPDFpolEIC-A} 
(left plot) or \texttt{NNPDFpolEIC-B} (right plot) fits,
while the dot-dashed lines refer to \texttt{NNPDFpol1.0}~\cite{Ball:2013lla}.}
\label{fig:chi2sets}
\end{figure}

The methodology for the determination of PDFs follows the one adopted in
Ref.~\cite{Ball:2013lla}, to which we refer for details. 
The only modifications are the following. First, we have re-tuned the genetic algorithm 
which is used for minimization, and the parameters which determine its stopping at the 
optimal fit.
This is required to obtain a good fit quality with EIC pseudodata, which are very accurate 
in comparison to their fixed-target counterparts and cover a wider kinematic region (see Fig.~\ref{fig:kinplot}).
In particular, we have used a larger population of mutants, increased the number of weighted 
training generations and tuned the stopping parameters.
Furthermore, we have redetermined the range in which preprocessing exponents are randomized, 
since the new information from EIC pseudodata may modify the large- and small-$x$ PDF behavior.
In Tab.~\ref{tab:preprocessing}, we show the values we use for the present fit, compared to \texttt{NNPDFpol1.0}.
We have checked that our choice of preprocessing exponents does not bias our fit, according 
to the procedure discussed in Sec.~4.1 of Ref.~\cite{Ball:2013lla}.
\begin{table}[t]
\small
\centering
\begin{tabular}{lcc|cc}
\toprule
& \multicolumn{2}{c|}{\texttt{NNPDFpol1.0}~\cite{Ball:2013lla}} 
& \multicolumn{2}{c}{\texttt{NNPDFpolEIC}}\\
PDF & $m$ & $n$ & $m$ & $n$\\
\midrule
$\Delta\Sigma(x,Q_0^2)$ & $[1.5, 3.5]$ & $[0.2, 0.7]$ 
                        & $[1.5, 3.5]$ & $[0.1, 0.7]$ \\
$\Delta g(x,Q_0^2)$     & $[2.5, 5.0]$ & $[0.4, 0.9]$ 
                        & $[2.0, 4.0]$ & $[0.1, 0.8]$ \\
$\Delta T_3(x,Q_0^2)$   & $[1.5, 3.5]$ & $[0.4, 0.7]$ 
                        & $[1.5, 3.0]$ & $[0.1, 0.6]$ \\
$\Delta T_8(x,Q_0^2)$   & $[1.5, 3.0]$ & $[0.1, 0.6]$ 
                        & $[1.5, 3.0]$ & $[0.1, 0.6]$ \\
\bottomrule
\end{tabular}
\caption{Ranges for the small- and large-$x$ 
preprocessing exponents.}
\label{tab:preprocessing}
\end{table}

Various general features of the  \texttt{NNPDFpolEIC-A} 
and \texttt{NNPDFpolEIC-B} PDF determinations  are summarized
in Tab.~\ref{tab:esttot}, compared to \texttt{NNPDFpol1.0}. 
These include the $\chi^2$ per data point of the final best-fit PDF compared to data,
(denoted as $\chi^2_{\mathrm{tot}}$), the  average and standard
deviation over the replica sample of the same figure of merit for each
PDF replica when compared to the corresponding data replica (denoted
as $\langle E\rangle \pm\sigma_E$) computed for the total, training
and validation sets, the average and standard deviation of the
$\chi^2$ of each replica when compared to data 
(denoted as $\langle\chi^{2(k)}\rangle$), 
and the average 
number of iterations of the genetic algorithm at stopping
$\langle \mathrm{TL}\rangle$ and its standard deviation over the
replica sample. A more detailed discussion of these quantities can 
be found in 
previous NNPDF papers, in particular in Refs.~\cite{Ball:2008by,Ball:2010de},
and Ref.~\cite{Ball:2013lla} for the polarized case.
\begin{table}[h]
\small
\centering
\begin{tabular}{cc|cc}
\toprule
& \texttt{NNPDFpol1.0}~\cite{Ball:2013lla}
& \texttt{NNPDFpolEIC-A} 
& \texttt{NNPDFpolEIC-B}\\
\midrule
$\chi^2_{\mathrm{tot}}$ 
& $0.77$  
& $0.79$
& $0.86$\\
$\langle E \rangle \pm \sigma_E$ 
& $1.82\pm 0.18$ 
& $2.24\pm 0.34$
& $2.44\pm 0.31$\\
$\langle E_{\mathrm{tr}} \rangle \pm \sigma_{E_{\mathrm{tr}}}$ 
& $1.66\pm 0.49$ 
& $1.87\pm 0.54$
& $1.81\pm 0.79$\\
$\langle E_{\mathrm{val}} \rangle \pm \sigma_{E_{\mathrm{val}}}$ 
& $1.88\pm 0.67$
& $2.61\pm 1.05$
& $2.47\pm 1.17$\\
$\langle\chi^{2(k)}\rangle \pm \sigma_{\chi^2}$
& $0.91\pm 0.12$
& $1.30\pm 0.31$
& $1.50\pm 0.30$\\
\midrule
$\langle \mathrm{TL} \rangle \pm \sigma_{\mathrm{TL}}$
& $6927\pm 3839$
& $7467\pm 3678$
& $19320\pm 14625$\\
\bottomrule
\end{tabular}
\caption{Statistical estimators and average training length for the 
\texttt{NNPDFpolEIC-A} and
\texttt{NNPDFpolEIC-B} with 
$N_{\mathrm{rep}}=100$ replicas, compared to the 
\texttt{NNPDFpol1.0} reference fit~\cite{Ball:2013lla}.}
\label{tab:esttot}
\end{table}

\begin{figure}[t]
\begin{center}
\epsfig{width=0.40\textwidth, figure=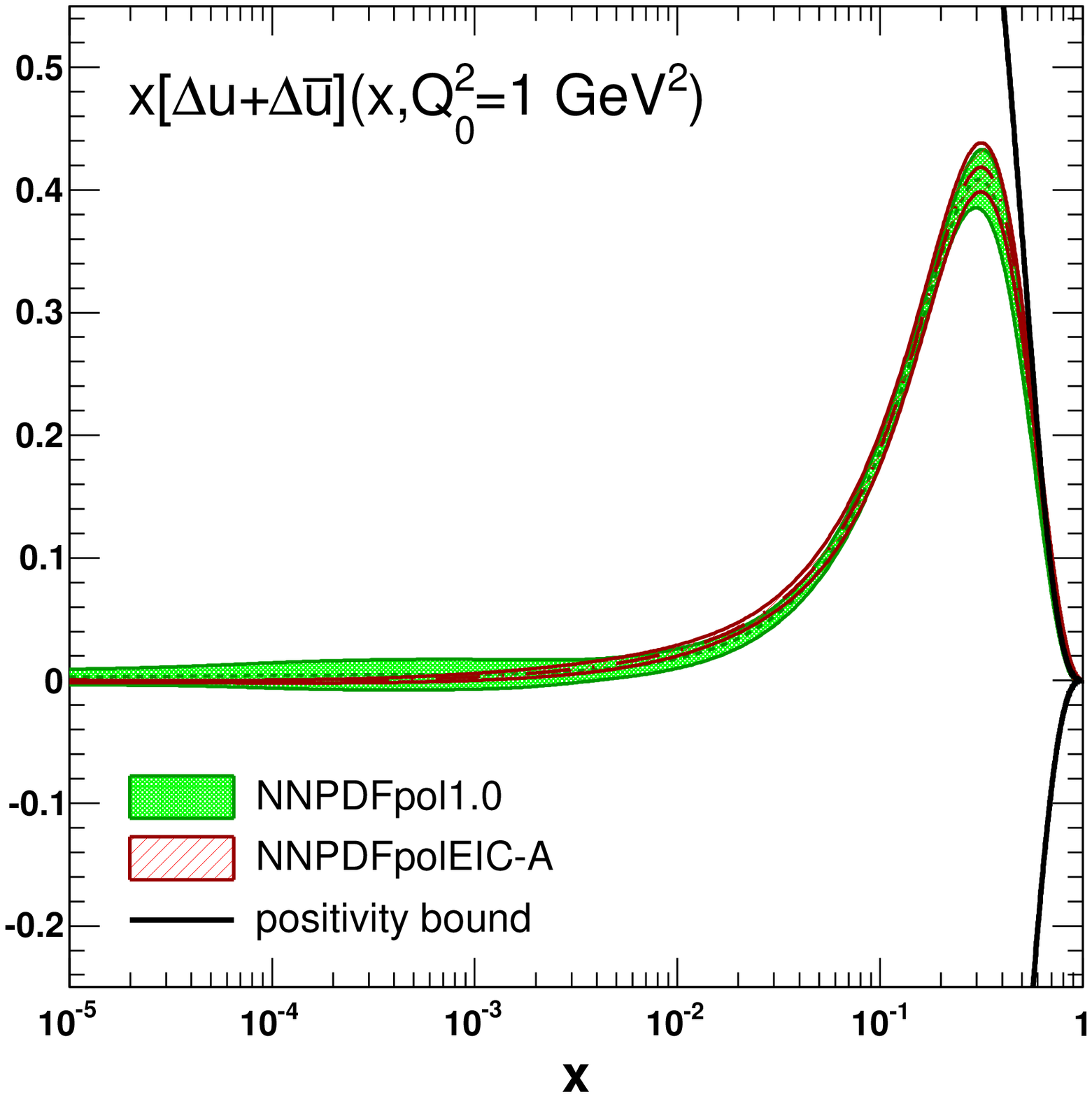}
\epsfig{width=0.40\textwidth, figure=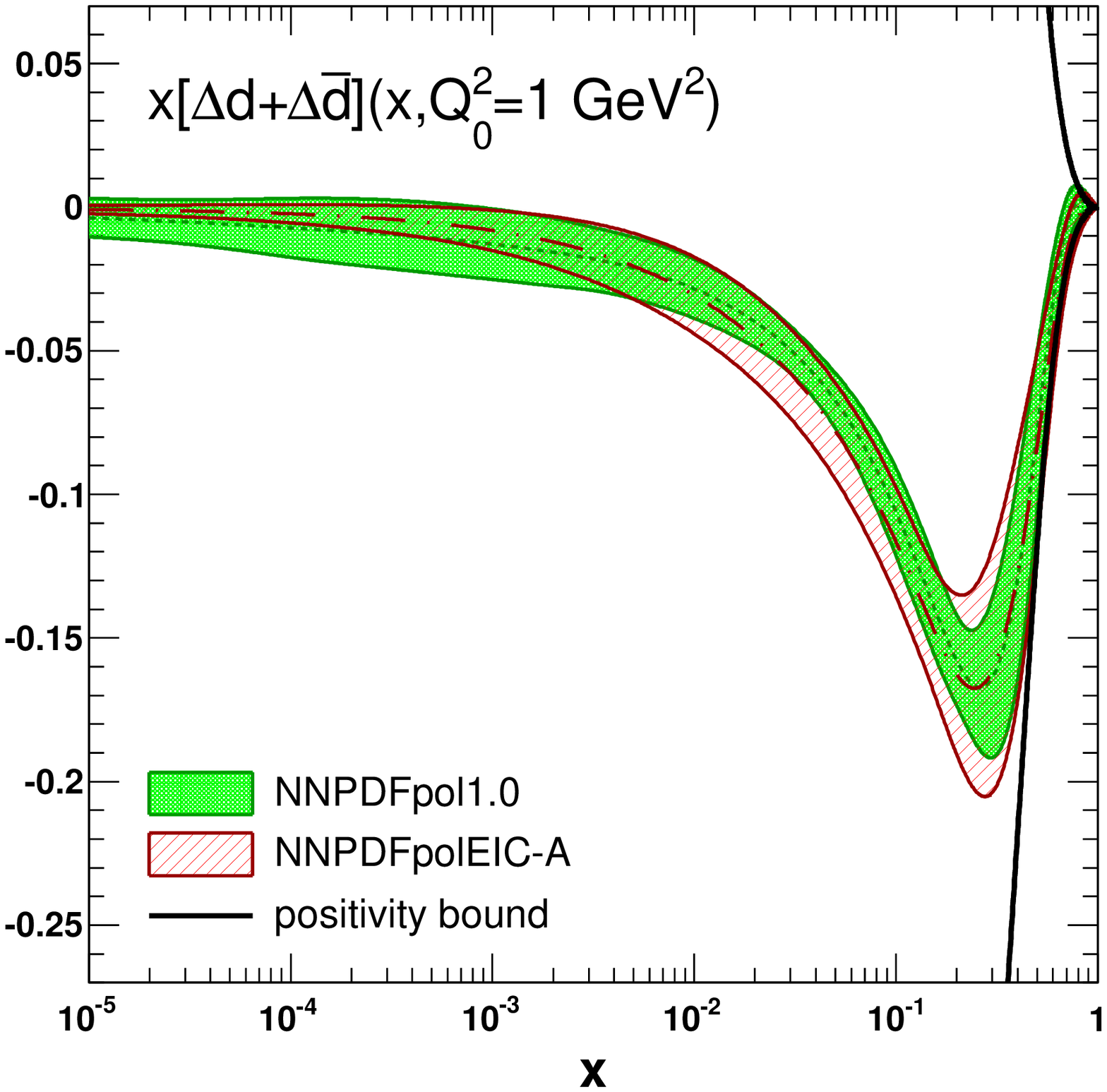}\\
\epsfig{width=0.40\textwidth, figure=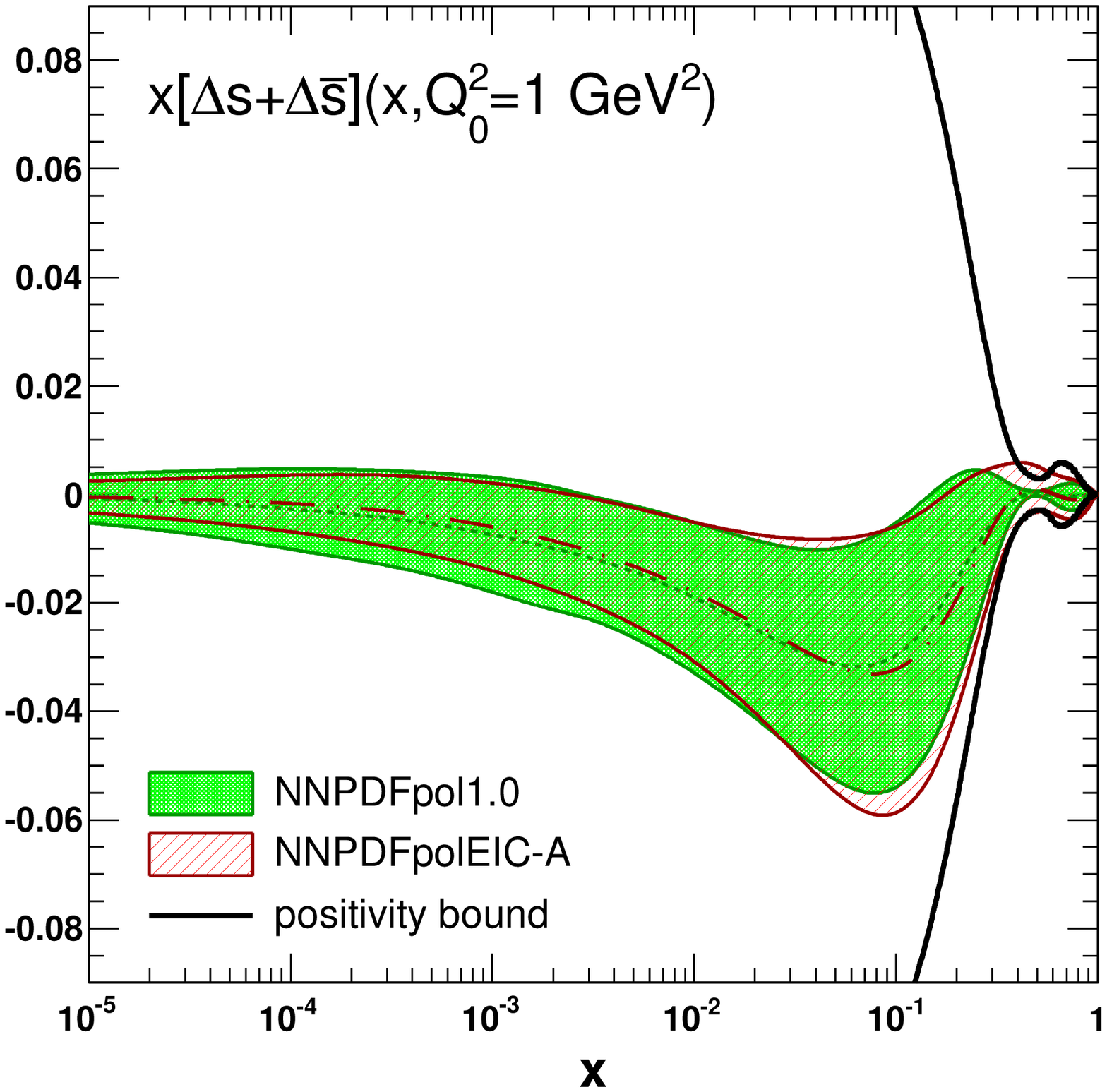}
\epsfig{width=0.40\textwidth, figure=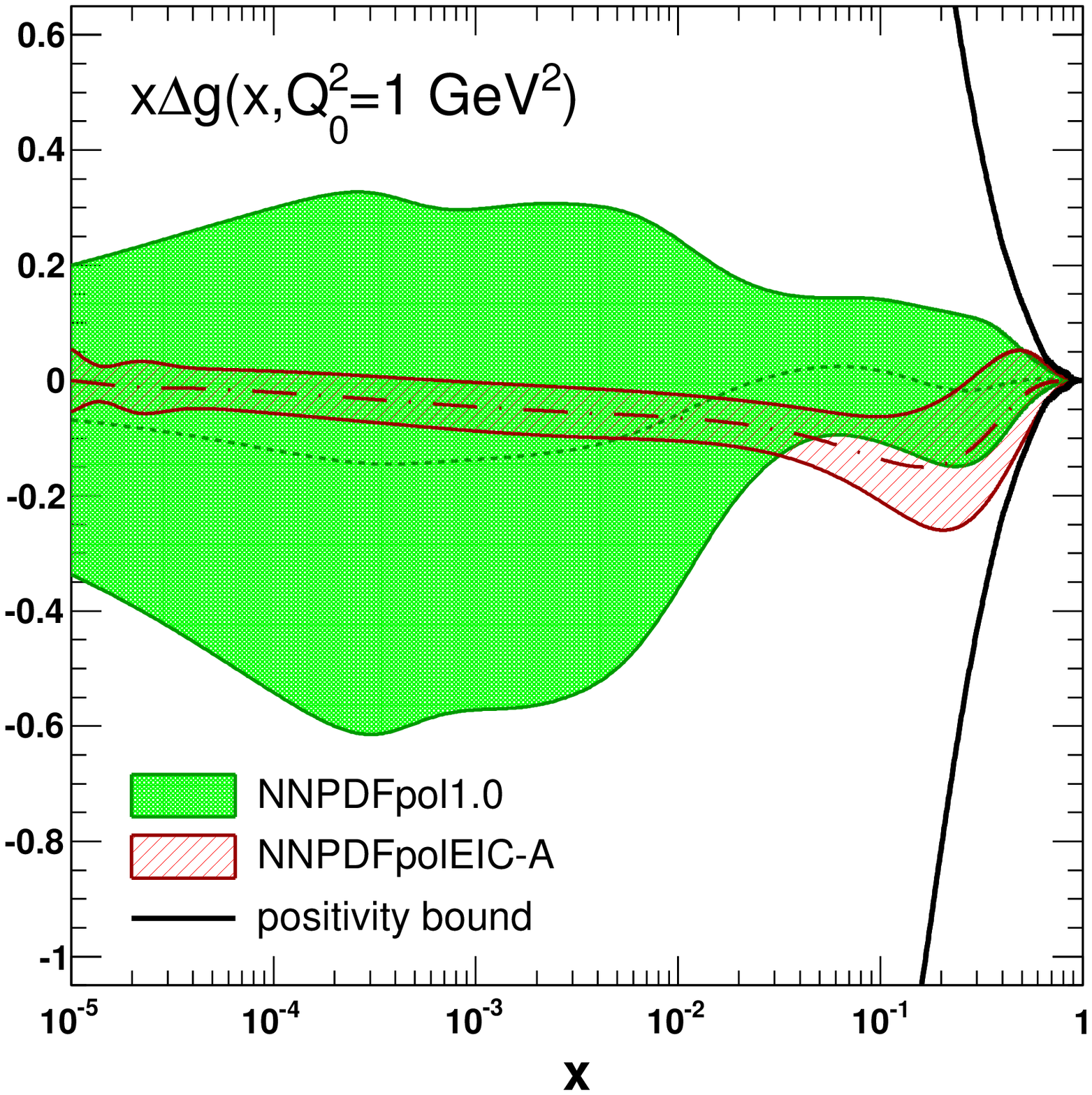}\\
\end{center}
\caption{The \texttt{NNPDFpolEIC-A} 
parton distributions at $Q_0^2=1$ GeV$^2$ plotted as a function of $x$ 
on a logarithmic scale, compared to \texttt{NNPDFpol1.0}~\cite{Ball:2013lla}.}
\label{fig:PDFsEIC1}
\end{figure}
\begin{figure}[t]
\begin{center}
\epsfig{width=0.40\textwidth, figure=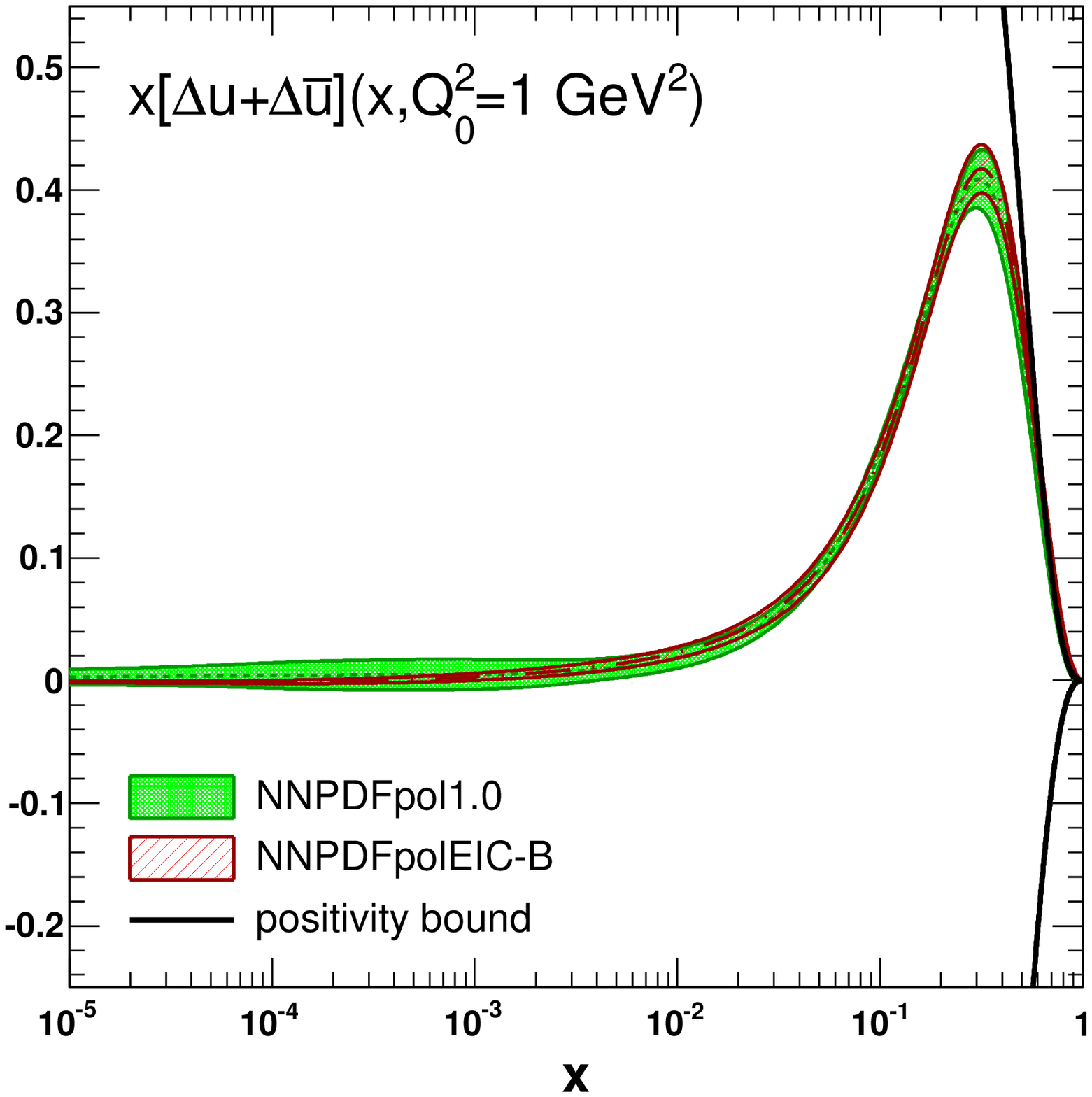}
\epsfig{width=0.40\textwidth, figure=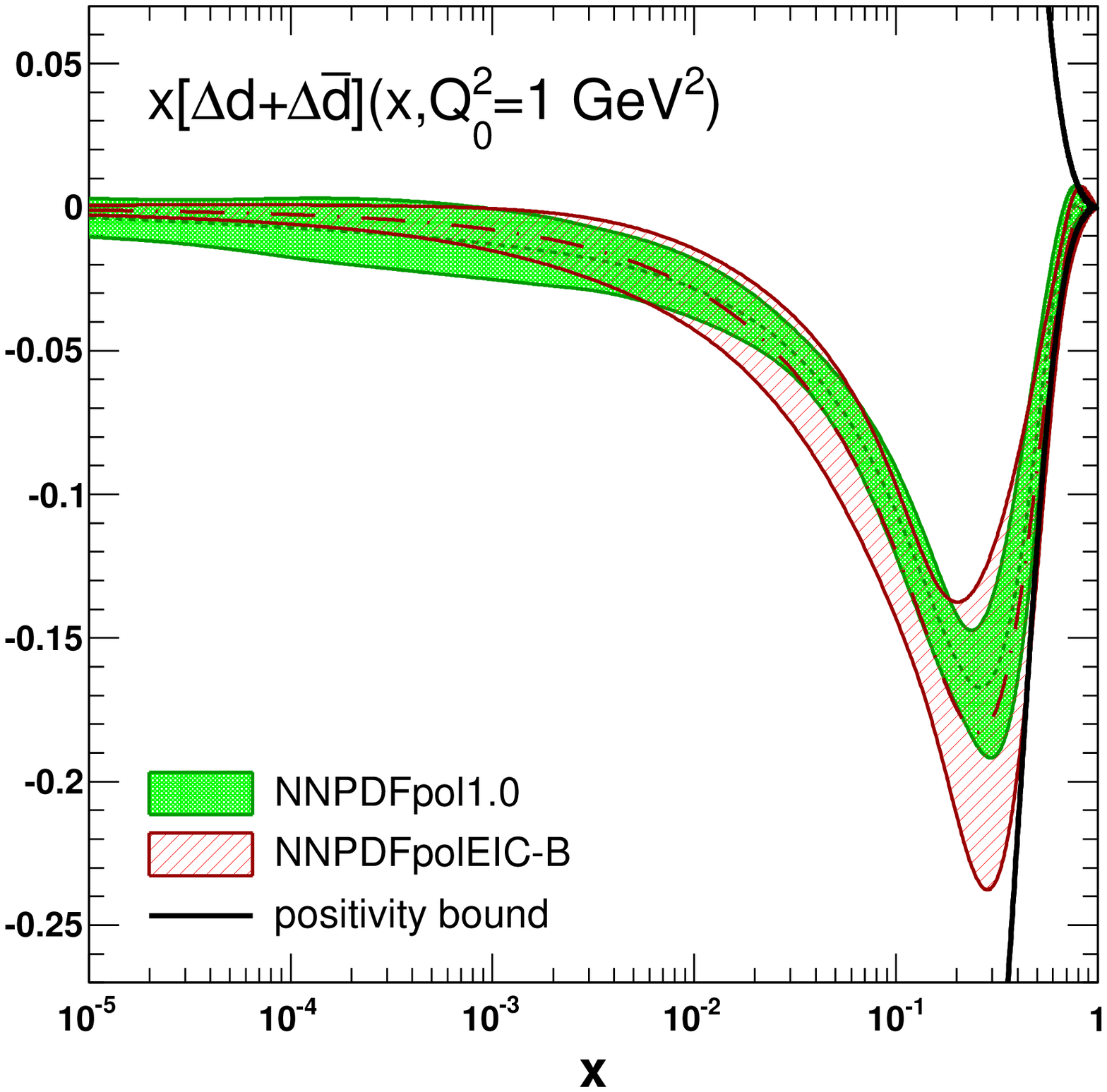}\\
\epsfig{width=0.40\textwidth, figure=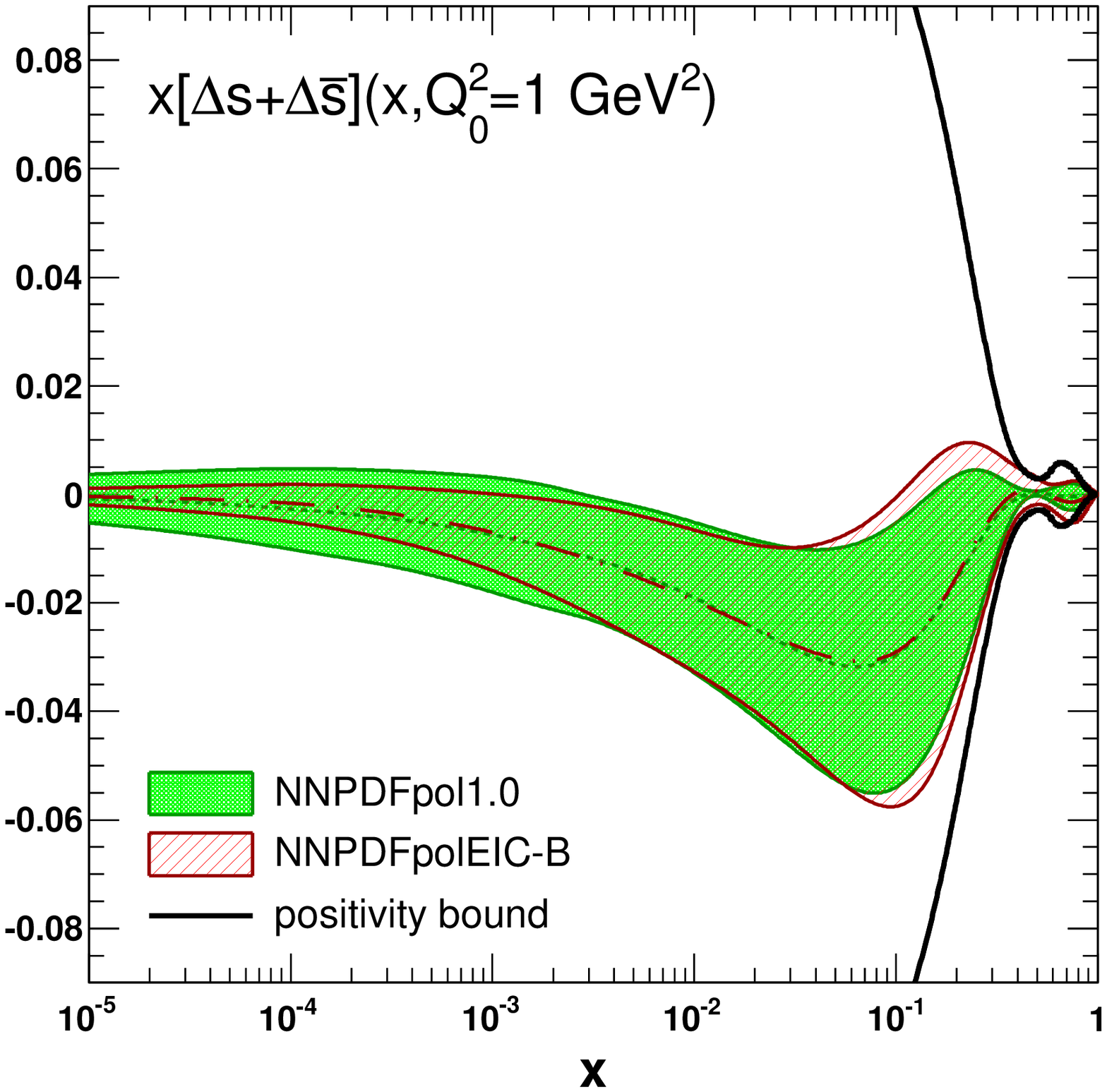}
\epsfig{width=0.40\textwidth, figure=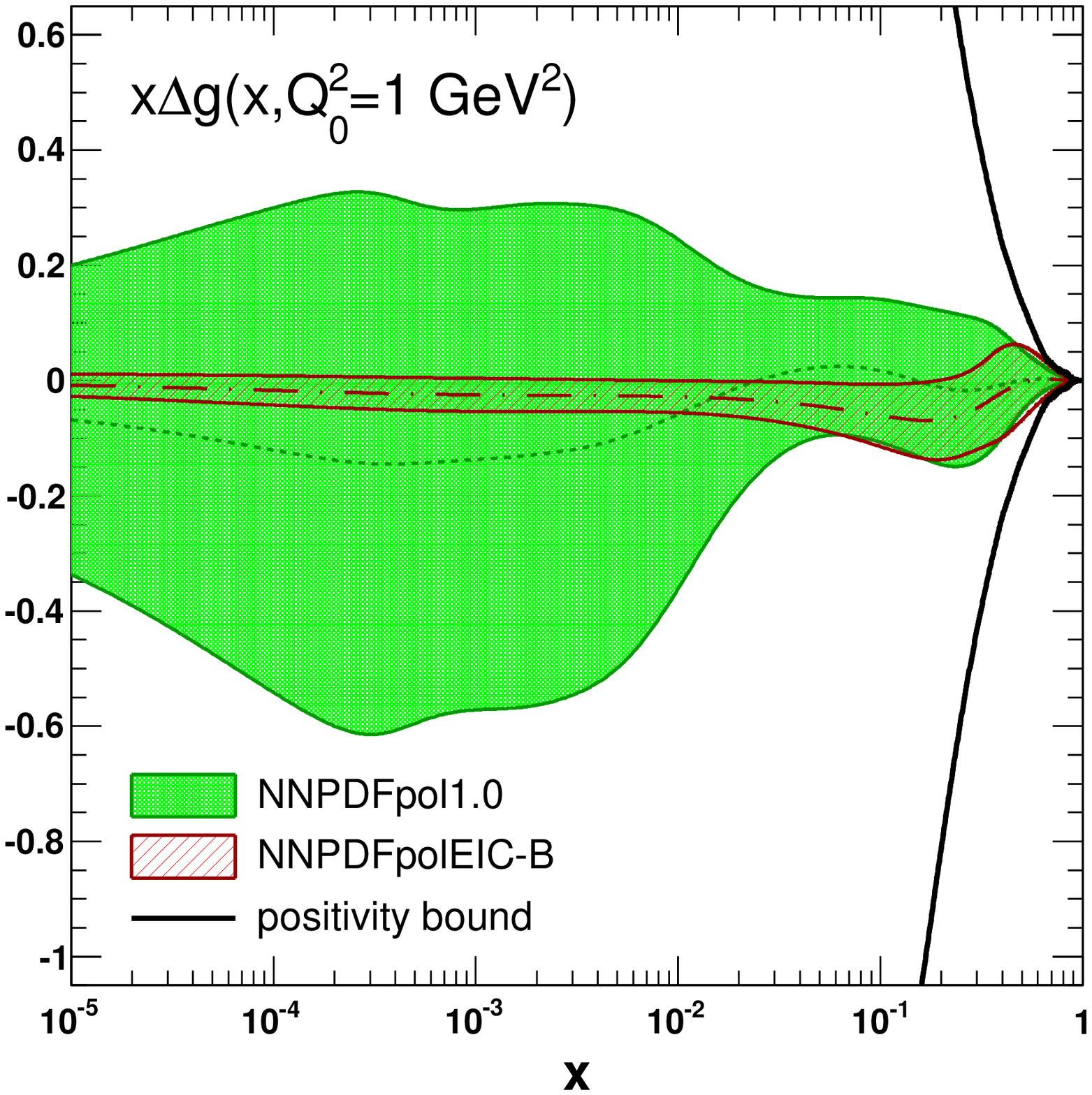}\\
\end{center}
\caption{Same as Fig.~\ref{fig:PDFsEIC1}, but for 
\texttt{NNPDFpolEIC-B}, compared to 
\texttt{NNPDFpol1.0}~\cite{Ball:2013lla}.}
\label{fig:PDFsEIC2}
\end{figure}
The  fit quality, as measured by $\chi^2_{\mathrm{tot}}$,
is comparable to that of \texttt{NNPDFpol1.0} 
($\chi^2_{\mathrm{tot}}=0.77$) for both the \texttt{NNPDFpolEIC-A} 
($\chi^2_{\mathrm{tot}}=0.79$) and the \texttt{NNPDFpolEIC-B}
($\chi^2_{\mathrm{tot}}=0.86$) fits. This shows that
our fitting procedure can easily accommodate EIC pseudodata.
The histogram of $\chi^2$ values for each data set included in our fits
is shown in Fig.~\ref{fig:chi2sets}, together with the
\texttt{NNPDFpol1.0}~\cite{Ball:2013lla} result;
the unweighted average 
$\langle\chi^2\rangle_{\mathrm{set}}\equiv\frac{1}{N_{\mathrm{set}}}\sum_{j=1}^{N_{\mathrm{set}}}\chi^2_{\mathrm{set,j}}$
and standard deviation over data sets are also shown.
As already pointed out in Ref.~\cite{Ball:2013lla}, $\chi^2$ values
significantly below one are found as a consequence of the fact that
information on correlated systematics is not available for most
experiments, and thus statistical and systematic errors are added in
quadrature. Note that this is not the case for the EIC pseudodata, for
which, as mentioned, no systematic uncertainty was included; this may
explain the somewhat larger (closer to one) value of the $\chi^2$ per
data point which is found when the pseudodata are included.

We notice that EIC pseudodata, which 
are expected to be rather more precise
than fixed-target DIS experimental data, require more training 
to be properly learned by the neural network. This is apparent in the
increase in $\langle TL\rangle$ in Tab.~\ref{tab:esttot}
when going from \texttt{NNPDFpol1.0} to \texttt{NNPDFpolEIC-A} and
then \texttt{NNPDFpolEIC-B}. 
We checked that the statistical features discussed above
do not improve if we run very long fits, 
up to $N_{\mathrm{gen}}^{\mathrm{max}}=50000$ generations,
without dynamical stopping. In particular we do not observe a 
decrease of the $\chi^2$ for those experiments whose value exceeds the average
by more than one sigma. This ensures that these deviations are not due to 
underlearning, \textit{i.e.} insufficiently long minimization.

Parton distributions from the
\texttt{NNPDFpolEIC-A} and \texttt{NNPDFpolEIC-B} 
fits are compared to \texttt{NNPDFpol1.0}~\cite{Ball:2013lla} in 
Figs.~\ref{fig:PDFsEIC1}-\ref{fig:PDFsEIC2} respectively.
In these plots, PDFs are displayed at 
$Q_0^2=1$ GeV$^2$ as a function of $x$ on a  logarithmic scale; all
uncertainties shown here are one-$\sigma$ bands. The 
positivity bound, obtained from the {\tt NNPDF2.3} NLO
unpolarized set~\cite{Ball:2012cx} as discussed in 
Ref.~\cite{Ball:2013lla}, is also drawn.

\begin{figure}[h]
\begin{center}
\epsfig{width=0.40\textwidth, figure=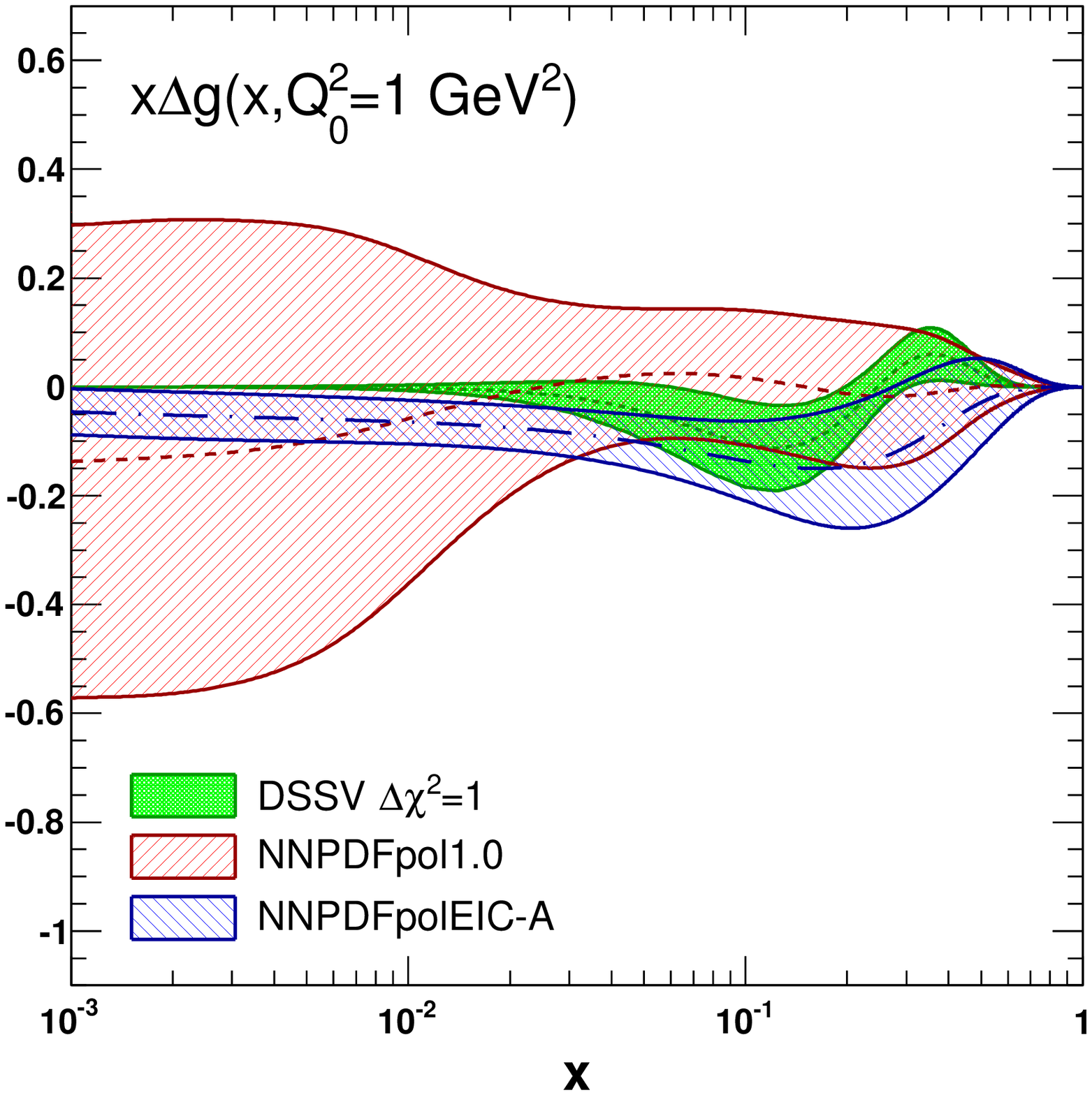}
\epsfig{width=0.40\textwidth, figure=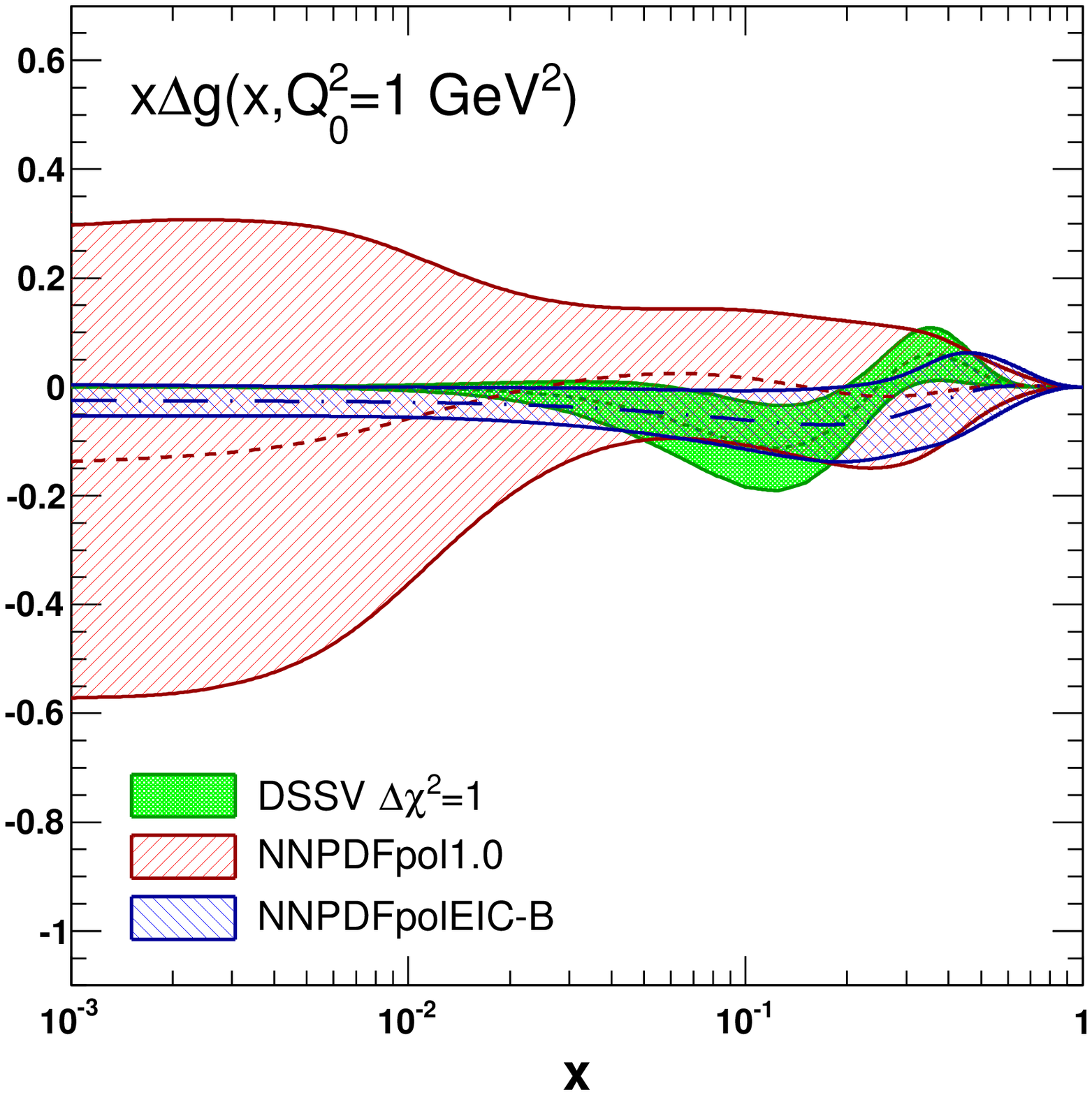}\\
\epsfig{width=0.40\textwidth, figure=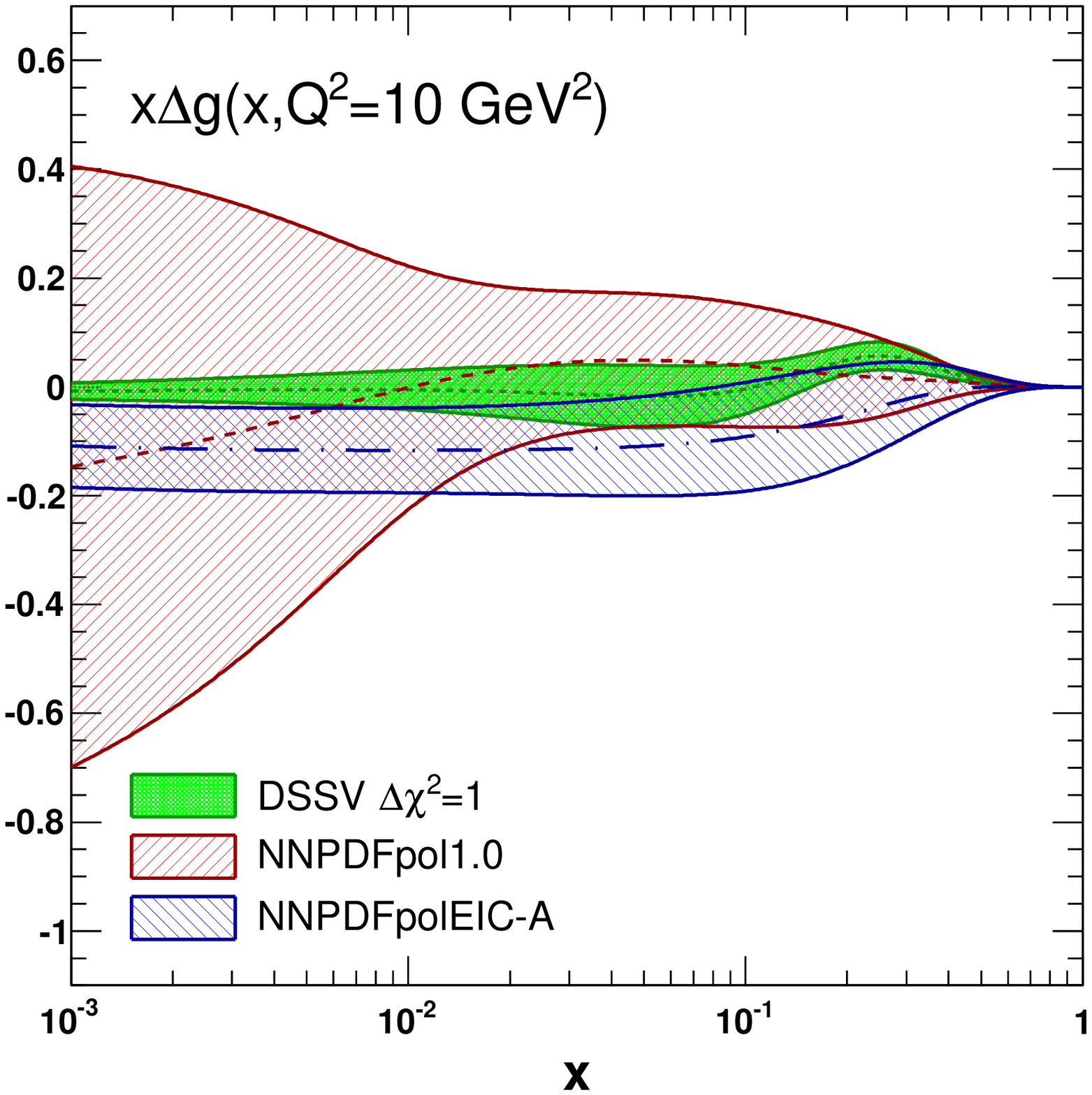}
\epsfig{width=0.40\textwidth, figure=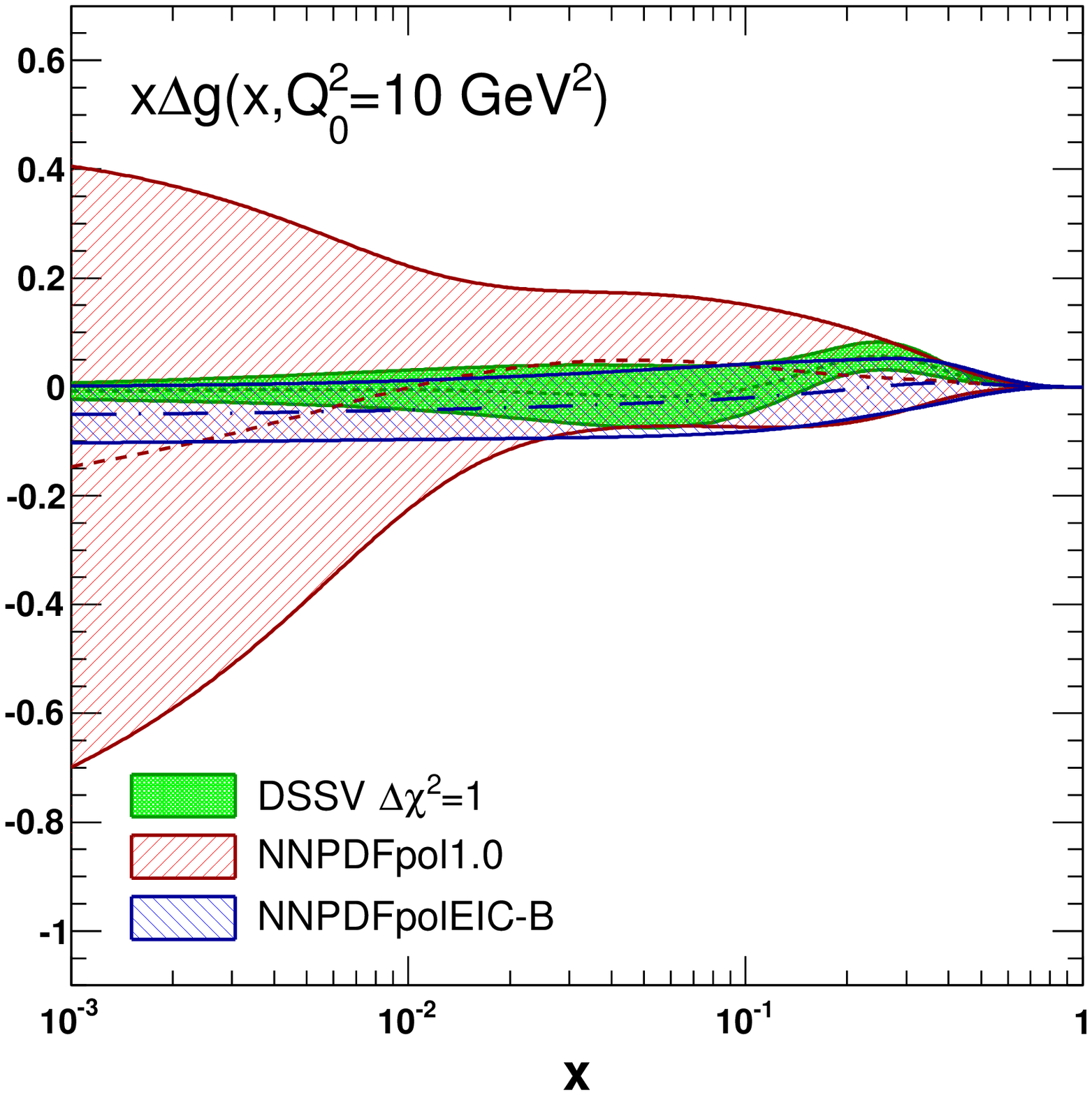}\\
\end{center}
\caption{The polarized gluon PDF $\Delta g(x,Q_0^2)$, at $Q_0^2=1$
GeV$^2$ (upper panels) and at $Q^2=10$ GeV$^2$ (lower panels),
in the \texttt{NNPDFpolEIC} PDF sets,
compared to DSSV~\cite{deFlorian:2009vb} and to
\texttt{NNPDFpol1.0}~\cite{Ball:2013lla}.}
\label{fig:gluon}
\end{figure}
\begin{figure}[h]
\begin{center}
\epsfig{width=0.40\textwidth, figure=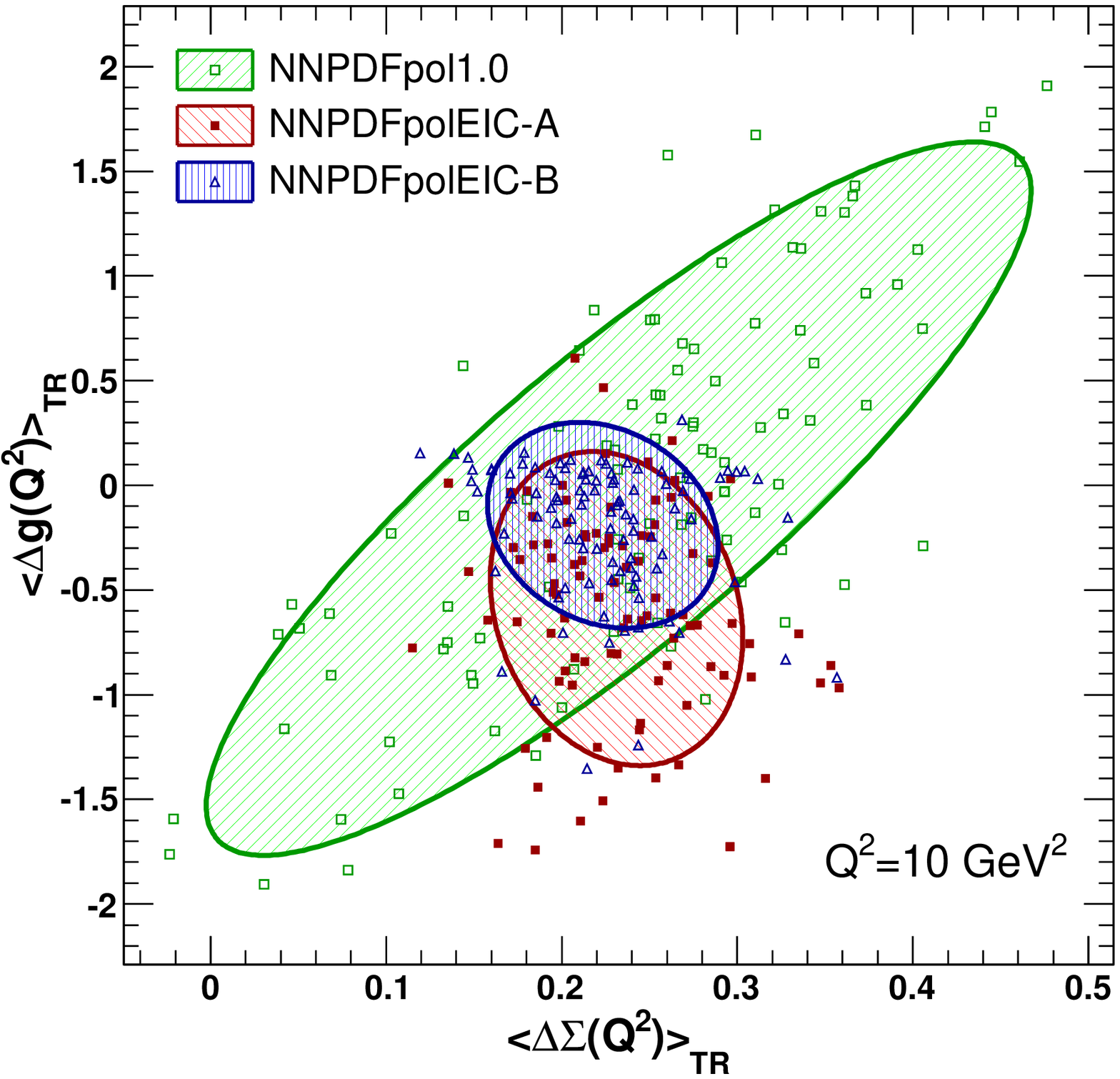}
\end{center}
\caption{One-$\sigma$ confidence region for the quark singlet and
  gluon first moments in the measured region,
  Eq.~(\ref{eq:trmoments}). The values for individual replicas are
  also shown.}
\label{fig:momcor}
\end{figure}
The most visible  impact of inclusive EIC pseudodata in
both our fits is the reduction of PDF uncertainties in the low-$x$ region
($x\lesssim 10^{-3}$) for  light flavors
and the gluon. 
The size of the effects is  different for 
different  PDFs. 
As expected, the most dramatic improvement is seen for the gluon,
while uncertainties on light quarks are only reduced by a significant
factor in the small $x$ region. The uncertainty on the strange
distribution is essentially unaffected:
unlike in Ref.~\cite{Aschenauer:2012ve}, we find  no
improvement on strangeness, due to the fact that we do not include
semi-inclusive kaon production data, contrary to what was done there.
When moving from \texttt{NNPDFpolEIC-A} to 
\texttt{NNPDFpolEIC-B} the gluon uncertainty decreases further, while
other PDF uncertainties are basically unchanged. 

In Fig.~\ref{fig:gluon} we compare the polarized gluon PDF 
in our  EIC fits to the 
DSSV~\cite{deFlorian:2009vb} and 
\texttt{NNPDFpol1.0}~\cite{Ball:2013lla}
parton determinations, both at $Q_0^2=1$~GeV$^2$ and  $Q^2=10$~GeV$^2$.
The DSSV uncertainty 
is the Hessian uncertainty computed assuming $\Delta\chi^2=1$, which
corresponds to the default uncertainty estimate in
Ref.~\cite{deFlorian:2009vb}. This choice may lead to somewhat
underestimated uncertainties: indeed, a more conservative uncertainty
estimate is also provided in
Ref.~\cite{deFlorian:2009vb}. Furthermore, it is known from
unpolarized
global PDF fits that a somewhat larger `tolerance' $T$ value
$\Delta\chi^2=T$~\cite{Pumplin:2002vw} should be adopted in order for
the distribution of $\chi^2$ values between different experiments in
the global fit to be reasonable (indeed this choice was made in
the polarized fit of Ref.~\cite{Hirai:2008aj}, with
$T=12.65$). 

It is clear that 
the gluon PDF from our fits including EIC pseudodata 
is approaching the DSSV PDF shape, especially at a lower scale where
the DSSV gluon does have some structure, despite the fact that at
higher scales, where much of the data is located, perturbative
evolution tends to wash out this shape. 
Also, this  is more pronounced as more EIC pseudodata
are included in our fit,
\textit{i.e.} moving from \texttt{NNPDFpolEIC-A} to
\texttt{NNPDFpolEIC-B}. 
This means that EIC data would be sufficiently accurate to reveal 
the polarized gluon structure, if any.

It is particularly interesting to examine how the EIC data affect the
determination of the  first moments 
\begin{equation}
\langle \Delta f(Q^2)\rangle\equiv\int_0^1dx\,\Delta f(x,Q^2)
\label{eq:fm}
\end{equation}
of the polarized PDFs $\Delta f(x,Q^2)$, as they are directly related
to the nucleon spin structure.
We have computed  the first moments, Eq.~(\ref{eq:fm}),
of the singlet, lightest quark-antiquark combinations and gluon for
the \texttt{NNPDFpolEIC-A} and \texttt{NNPDFpolEIC-B} PDF sets.
The corresponding central values  and one-$\sigma$ uncertainties
at $Q_0^2=1$ GeV$^2$ are shown in Tab.~\ref{tab:fullmom}, compared to 
\texttt{NNPDFpol1.0}~\cite{Ball:2013lla}. 
\begin{table}[t]
\small
\centering
\begin{tabular}{lccccc}
\toprule
Fit & $\langle\Delta\Sigma\rangle$ 
& $\langle\Delta u +\Delta\bar{u}\rangle$ 
& $\langle\Delta d +\Delta\bar{d}\rangle$ 
& $\langle\Delta s +\Delta\bar{s}\rangle$ 
& $\langle\Delta g \rangle$\\
\midrule
\texttt{NNPDFpol1.0}~\cite{Ball:2013lla}
&  $0.22\pm 0.20$
&  $0.80\pm 0.08$
& $-0.46\pm 0.08$
& $-0.13\pm 0.10$
& $-1.15 \pm 4.19$\\
\midrule
\texttt{NNPDFpolEIC-A}
&  $0.24\pm 0.08$
&  $0.82\pm 0.02$
& $-0.45\pm 0.02$
& $-0.13\pm 0.07$ 
& $-0.59 \pm 0.86$\\
\texttt{NNPDFpolEIC-B}
&  $0.21\pm 0.06$
&  $0.81\pm 0.02$
& $-0.47\pm 0.02$
& $-0.12\pm 0.07$
& $-0.33 \pm 0.43$\\
\bottomrule
\end{tabular}
\caption{First moments of the polarized quark distributions at 
$Q_0^2=1$ GeV$^2$ for the fits in the present analysis, compared to 
\texttt{NNPDFpol1.0}~\cite{Ball:2013lla}.}
\label{tab:fullmom}
\end{table}

It is clear that EIC pseudodata 
reduce all   uncertainties significantly.
Note that
moving from \texttt{NNPDFpolEIC-A} to 
\texttt{NNPDFpolEIC-B} does not improve significantly
the uncertainty on quark-antiquark first moments, but it reduces the
uncertainty on the gluon first moment by a factor two.
However, it is worth noticing that, despite a reduction of the
uncertainty on  the 
gluon first moment, even for the most accurate
\texttt{NNPDFpolEIC-B} fit, the value remains compatible with zero
even though the central value is sizable (and negative).

In order to assess the residual extrapolation uncertainty 
on the singlet and gluon first moments, we determine the contribution
to them from the data range $x\in[10^{-3},1]$, \textit{i.e.} 
\begin{equation}
\langle\Delta\Sigma(Q^2)\rangle_{\mathrm{TR}}
\equiv
\int_{10^{-3}}^{1}dx\,\Delta\Sigma(x,Q^2)
\mbox{ ,}
\ \ \ \ \ \ \ \ \ \ 
\langle\Delta g(Q^2)\rangle_{\mathrm{TR}}
\equiv
\int_{10^{-3}}^{1}dx\,\Delta g(x,Q^2)
\mbox{ .}
\label{eq:trmoments}
\end{equation} 
The first moments Eq.~(\ref{eq:trmoments}) are given in
Tab.~\ref{tab:trmomenta} at
$Q_0^2=1$ GeV$^2$ 
and $Q^2=10$ GeV$^2$, where results for central values, uncertainties, and
correlation coefficients between the gluon and quark are collected.

Comparing the results at $Q^2=1$ GeV$^2$ of Tab.~\ref{tab:fullmom}
and Tab.~\ref{tab:trmomenta} we see that in the  \texttt{NNPDFpol1.0}
PDF determination for the quark singlet combination the uncertainty on 
the full first moment is about twice as large as that from the measured 
region, and for the gluon it is about four times as large. 
The difference is due to the extra uncertainty coming from the extrapolation. 
In \texttt{NNPDFpolEIC-B} the corresponding increases are by 20\% for the 
quark and 30\% for the gluon, which shows that thanks to EIC data the extrapolation
uncertainties would be largely under control. 
The correlation coefficient $\rho$ significantly decreases upon inclusion of the EIC
data: this means that the extra information contained in these data allows for an 
independent determination of the quark and gluon first moments.

\begin{table}[t]
\small
\centering
\begin{tabular}{lcc|ccc}
\toprule
& \multicolumn{2}{c|}{$Q^2=1$ GeV$^2$}
& \multicolumn{3}{c}{$Q^2=10$ GeV$^2$}\\
& $\langle\Delta\Sigma(Q^2)\rangle_{\mathrm{TR}}$ 
& $\langle\Delta g(Q^2)\rangle_{\mathrm{TR}}$
& $\langle\Delta\Sigma(Q^2)\rangle_{\mathrm{TR}}$ 
& $\langle\Delta g(Q^2)\rangle_{\mathrm{TR}}$
& $\rho(Q^2)$ \\
\midrule
\texttt{NNPDFpol1.0}~\cite{Ball:2013lla}
&  $0.25\pm 0.09$
& $-0.26\pm 1.19$ 
&  $0.23\pm 0.16$ 
& $-0.06\pm 1.12$ 
& $+0.861$\\ 
\midrule
\texttt{NNPDFpolEIC-A}
&  $0.27\pm 0.06$
& $-0.53\pm 0.37$ 
&  $0.23\pm 0.05$ 
& $-0.59\pm 0.50$ 
& $-0.186$\\ 
\texttt{NNPDFpolEIC-B}
&  $0.24\pm 0.05$
& $-0.23\pm 0.25$ 
&  $0.22\pm 0.04$ 
& $-0.19\pm 0.32$ 
& $-0.103$\\
\bottomrule
\end{tabular}
\caption{The singlet and gluon truncated first moments and
their one-$\sigma$ uncertainties at $Q^2=1$ GeV$^2$ and $Q^2=10$ GeV$^2$ 
for the \texttt{NNPDFpolEIC} PDF sets,
compared to \texttt{NNPDFpol1.0}~\cite{Ball:2013lla}.
The correlation coefficient $\rho$ at $Q^2=10$ GeV$^2$ 
is also provided.}
\label{tab:trmomenta}
\end{table}

In Fig.~\ref{fig:momcor}, we plot the one-$\sigma$ 
confidence region in the 
$(\langle\Delta\Sigma(Q^2)\rangle_{\mathrm{TR}},
\langle\Delta g(Q^2)\rangle_{\mathrm{TR}})$ plane
at $Q^2=10$ GeV$^2$, for 
\texttt{NNPDFpolEIC-A},
\texttt{NNPDFpolEIC-B} and \texttt{NNPDFpol1.0}~\cite{Ball:2013lla}.
The main result of our analysis, Fig.~\ref{fig:momcor}, can be
directly compared to Fig.~8 of Ref.~\cite{Aschenauer:2012ve}, which was
based on the DSSV fit and is comparable to our \texttt{NNPDFpolEIC-B}
results.  In 
both analyses EIC pseudodata determine the
singlet  first moment in the measured region with an uncertainty of 
about $\pm 0.05$. 

On the other hand, in Ref.~\cite{As chenauer:2012ve} the uncertainty
on the gluon  was found to be about $\pm 0.02$, while we get a much larger
result of $\pm 0.30$.  One may wonder whether this difference may be
due at least in part to the fact that the DSSV fit on which the result of
Ref.~\cite{Aschenauer:2012ve} is based also includes jet production
and pion production data from RHIC, which may reduce the gluon
uncertainty. To answer this, we have computed the contribution to the
gluon first moment (again at  $Q^2=10$ GeV$^2$) from
the reduced region $0.05\le x\le 0.2$, where the RHIC data are
located. We find that the uncertainty on the contribution to the
gluon first moment in this restricted range is $\pm 0.083$ using
\texttt{NNPDFpolEIC-B}, while it is  $\pm 0.147$  with 
\texttt{NNPDFpol1.0} and ${}^{+0.129}_{-0.164}$ with
DSSV~\cite{Aschenauer:2013woa}.  We
conclude that before the EIC data are added, the uncertainties
in \texttt{NNPDFpol1.0} and DSSV are quite
similar despite the fact that DSSV also includes RHIC data.
Hence, the larger gluon uncertainty we find for the
\texttt{NNPDFpolEIC-B} fit in comparison to
Ref.~\cite{Aschenauer:2012ve} is likely to be due to
our  more 
flexible PDF parametrization, though some difference might also come from
the fact that the SIDIS pseudodata included in 
Ref.~\cite{Aschenauer:2012ve} provide additional information 
on the gluon through scaling violations of the fragmentation 
structure function $g_1^h$ (of course this also introduces an uncertainty
related to the fragmentation functions which is difficult to quantify).

In summary, the EIC data would entail a very
considerable reduction in the uncertainty on the polarized gluon.
 They
would provide first evidence for a possible nontrivial $x$ shape of
the polarized gluon distribution. They would also provide evidence for
or against a possible large gluon contribution to the nucleon spin,
though the latter goal would still be reached with a sizable residual
uncertainty. 
Additional measurements at an EIC, such as the charm polarized
structure function, $g_1^c$, might provide more information on $\Delta
g$ and its first moment.

\bigskip
\bigskip

{\bf\noindent  Acknowledgments \\}
We would like to thank M.~Stratmann for suggesting us to
work on this project and for providing us with the 
EIC pseudodata
from Ref.~\cite{Aschenauer:2012ve}. 
The research of J.~R. has been partially supported by a Marie Curie 
Intra-European Fellowship of the European Community's 7th
Framework Programme under contract number PIEF-GA-2010-272515. 
S.~F., E.~R.~N. and G.~R. are partly supported by a PRIN2010 grant.

\end{document}